\documentclass[onecolumn,english,reprint, longbibliography, superscriptaddress, breaklinks=true, showkeys, showpacs=false, nofootinbib]{revtex4}
\usepackage[T1]{fontenc}
\usepackage[latin9]{inputenc}
\usepackage{color,footnote}
\usepackage[english]{babel}
\usepackage{amsmath,amssymb}
\usepackage{mathtools,bbm}
\usepackage{subfigure}
\usepackage{graphicx}
\usepackage{physics,braket}
\usepackage{comment}
\usepackage{ragged2e}
\usepackage{mathtools}
\usepackage[varg]{txfonts} 
\usepackage[unicode=true,pdfusetitle,bookmarks=true,bookmarksnumbered=false,bookmarksopen=false,breaklinks=true,pdfborder={0 0 0},backref=false,colorlinks=true]
 {hyperref}
\makeatletter
\@ifundefined{textcolor}{}
{%
 \definecolor{BLACK}{gray}{0}
 \definecolor{WHITE}{gray}{1}
 \definecolor{RED}{rgb}{1,0,0}
 \definecolor{GREEN}{rgb}{0,1,0}
 \definecolor{BLUE}{rgb}{0,0,1}
 \definecolor{CYAN}{cmyk}{1,0,0,0}
 \definecolor{MAGENTA}{cmyk}{0,1,0,0}
 \definecolor{YELLOW}{cmyk}{0,0,1,0}
}
\pdfoutput=1
\hypersetup{colorlinks=true,citecolor=blue,linkcolor=cyan,urlcolor=blue,filecolor= green, breaklinks=true}
\usepackage{url}
\usepackage{breakurl}
\makeatother

\newcommand{\mf}{\mathfrak}

\begin{document}

\author{Diego S. Starke\href{https://orcid.org/0000-0002-6074-4488}{\includegraphics[scale=0.05]{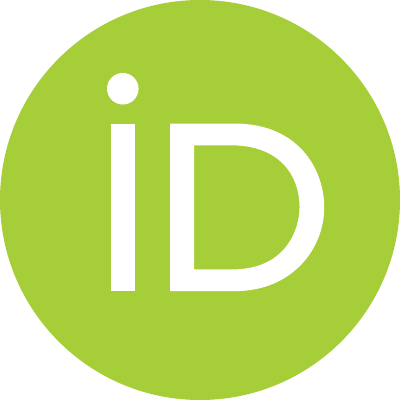}}}
\email[]{starkediego@gmail.com}
\affiliation{Department of Physics, Center for Natural and Exact Sciences, Federal University of Santa Maria, Roraima Avenue 1000, Santa Maria, Rio Grande do Sul, 97105-900, Brazil}

\author{Marcos L. W. Basso\href{https://orcid.org/0000-0001-5456-7772}{\includegraphics[scale=0.05]{orcidid.pdf}}}
\email[]{marcoslwbasso@hotmail.com}
\affiliation{Center for Natural and Human Sciences, Federal University of ABC, States Avenue 5001, Santo Andr\'e, S\~ao Paulo, 09210-580, Brazil}

\author{Jonas Maziero\href{https://orcid.org/0000-0002-2872-986X}{\includegraphics[scale=0.05]{orcidid.pdf}}}
\email[]{jonas.maziero@ufsm.br}
\affiliation{Department of Physics, Center for Natural and Exact Sciences, Federal University of Santa Maria, Roraima Avenue 1000, Santa Maria, Rio Grande do Sul, 97105-900, Brazil}

\selectlanguage{english}

\title{An Updated Quantum Complementarity Principle}

\begin{abstract}
Bohr's complementarity principle has long been a fundamental concept in quantum mechanics, positing that, within a given experimental setup, a quantum system (or quanton) can exhibit either its wave-like character, denoted as $W$, or its particle-like character, denoted as $P$, but not both simultaneously. Modern interpretations of Bohr's complementarity principle acknowledge the coexistence of these aspects in the same experiment while introducing the constraint $W + P \le \alpha$. 
Notably, estimations of $W$ or $P$ frequently rely on indirect retrodiction methods, a practice that has led to the claim of the violation of Bohr's complementarity principle. By taking a different route, recent advancements demonstrate that quantum complementarity relations can be rigorously derived from the axioms of quantum mechanics. To reconcile these observations and eliminate potential paradoxes or violations, we propose an updated formulation for the quantum complementarity principle, which is stated as follows: \textit{For a given quantum state preparation $\rho_t$ at a specific instant of time $t$, the wave and particle behaviors of a quanton are constrained by a complementarity relation $\mf{W}(\rho_t) + \mf{P}(\rho_t) \le \alpha(d)$, which is derived directly from the axioms of quantum mechanics.}
\end{abstract}


\date{\today}

\maketitle

\section{Introduction}

At the 1927 International Congress of Physics in Como, Niels Bohr introduced his groundbreaking complementarity principle \cite{Bohr1928,Bohr1958,Plotnitsky,khrennikov}, which has since become a cornerstone of quantum mechanics. According to Bohr's qualitative statement, quantum systems (or quantons) exhibit properties that are equally real but mutually exclusive within a given experimental setup, with the wave-particle duality serving as its most prominent illustration.

However, it was not until 1979 that the first quantitative formulation of Bohr's principle was put forward by Wootters and Zurek~\cite{Wootters1979}, rooted in an information-theoretic inequality. As succinctly stated by these authors, ``the more clearly we wish to observe the wave nature of light, the more information we must give up about its particle properties''.

Nearly a decade later, in 1988, Greenberger and Yasin~\cite{Greenberger1988} synthesized these two facets of quantons into a quantitative complementarity relation like $W + P \le \alpha$ and expressed as
\begin{equation}
\mathcal{V}^2 + \mathcal{P}^2 \le 1. 
\label{eq:cr}
\end{equation}
Here, $W=\mathcal{V}^2$ represents a measure of wave-like behavior, $P=\mathcal{P}^2$ represents a measure of particle-like behavior, and $\alpha=1$. This inequality, applicable to a given experiment, encapsulates the idea that both wave and particle aspects can be observed in the same experimental setup; however, there exists an inherent limitation on the extent to which they can manifest. As introduced by Greenberger and Yasin, the interferometric visibility $\mathcal{V}$ is defined as
\begin{equation}
\mathcal{V} \equiv \frac{p^{\max}_i - p^{\min}_i}{p^{\max}_i + p^{\min}_i}.
\label{eq:vis}
\end{equation}
This definition is applied to a Mach--Zehnder interferometer (MZI), with the index $i = 1,2$ representing the detectors used to measure the maximum and minimum probabilities $p_i$ when varying the phase within the phase shifter~\cite{Greenberger1988,Starke2023}.

For the particle-like measure, Greenberger and Yasin introduced the so called predictability $\mathcal{P} \equiv \abs{w_1 - w_2}$, where $w_j$ is the probability that a quanton follows the path $j$ inside the interferometer. The predictability represents a measure of \textit{a priori} which-way information once it is related to the probability of correctly guessing the path of the quanton within the MZI. Notably, the visibility $\mathcal{V}$ is intended to indirectly quantify wave-like behavior within the MZI through probabilities obtained using detectors positioned outside the MZI. Therefore, it is important to acknowledge that the predictability and the visibility depend on probabilities estimated at different instants of time.

Subsequently, Englert~\cite{Englert1996} extended the concept of a complementarity relation akin to Eq.~\eqref{eq:cr} using $\mathcal{V}$ as the measure of wave-like behavior and the distinguishability $D$ as the measure of particle-like behavior. Distinguishability quantifies the which-way information that is available through the path detectors inserted into the arms of the MZI, effectively representing \textit{a posteriori} path information. Notably, Englert's work expressed these wave-particle measures as functions of some state preparations of the two-level quanton.

Over recent decades, there has been a growing interest in quantifying Bohr's complementarity principle through complementarity relations. The pioneer work by D\"urr~\cite{Durr2001} and Englert et al.~\cite{Englert2008} laid the foundation for establishing reasonable properties that any measure of wave-like and particle-like behavior should satisfy. A significant development has been the formulation of complementarity relations for $d$-dimensional quantons, commonly referred to as qudits. This became possible as researchers recognized that quantum coherence~\cite{Baumgratz2014} provides a natural extension of the wave-like aspect of a qudit~\cite{Bera2015,Bagan2016,Mishra2019,Starke2023} and captures the essence of the wave-like behavior, i.e., the path superposition inside the interferometer~\cite{Wheeler1984}. 
Furthermore, a multitude of complementarity relations have emerged to account for the state preparation of the quantum system~\cite{Coles2014,Bera2015,Bagan2016,Angelo2015,Coles2016,Bagan2018,Qureshi2019,Basso2020}. These complementarity relations are generally expressed in the form $\mf{W}(\rho_t)+\mf{P}(\rho_t)\le\alpha(d)$, where $\alpha(d)$ is a constant that depends on the dimension $d$ of the quanton and $\rho_t$ is the density matrix associated with the system at time $t$. In this work, although it is always possible, we choose not to normalize these functions to make the dimension of the system explicit.

Motivated by these valuable contributions that have expanded the significance and generality of complementarity relations, a pioneering step forward was taken by introducing a procedure for deriving quantum complementarity relations that quantitatively describe the complementarity principle based on the fundamental postulates of quantum mechanics \cite{Basso2020} and which satisfy the basic properties established in Refs.~\cite{Durr2001,Englert2008}. This new approach explicitly utilizes the quantum state postulate. Built upon recent developments, we propose to review the formulation of the quantum complementarity principle in light of these quantum complementarity relations. Our reformulation implies that the quantum complementarity principle becomes an integral part of quantum mechanics, rather than a concept existing alongside it. Moreover, our proposal effectively tackles potential breaches of Bohr's complementarity principle, as the Afshar's experiment~\cite{Afshar2005,Afshar2007}, and resolves peculiar situations, as the Wheeler's delayed choice experiment~\cite{Wheeler1984,Ma2016}.

The remainder of this article is organized as follows. In Sec.~\ref{sec:update}, we review the procedure established in Ref.~\cite{Basso2020} in order to derive quantum complementarity relations from the axioms of quantum mechanics and propose our reformulation of the quantum complementarity principle.  Sec.~\ref{sec:intexp} is reserved for the discussion of the different interferometric experiments, such as Afshar's experiment and Wheeler's delayed choice experiment, as well as two other experimental setups that highlight the necessity and importance of our reformulation of the quantum complementarity principle. Finally, in Sec.~\ref{sec:conc}, we give our conclusions.

\section{An Updated Quantum Complementarity Principle}
\label{sec:update}

Let us briefly recall the set of density matrices acting on a Hilbert space $\mathcal{H}$, i.e., $\mathcal{D}(\mathcal{H}) \equiv \{\rho_t \in \mathcal{L}(\mathcal{H}) \ | \ \rho_t \ge 0, \ \Tr\rho_t = 1 \}$ where $\mathcal{L}(\mathcal{H})$ is the set of linear operators on $\mathcal{H}$. According to the quantum state postulate, for every quantum system associated with a Hilbert space $\mathcal{H}$, the quantum state of the system at any given time is described by a density matrix $\rho_t \in \mathcal{D}(\mathcal{H})$.

The procedure established in Ref.~\cite{Basso2020} starts with the selection of a wave-like measure that aligns with the fundamental properties mentioned above \cite{Durr2001,Englert2008}, exemplified by any quantum coherence measure \cite{Baumgratz2014}. For instance, let us consider the $l_1-$norm quantum coherence  given by $C_{l_{1}}(\rho_t) = \sum_{j\ne k}|\rho_{jk}|$, where $\rho_{jk}$ are the elements of the density operator $\rho_t$ on a given basis (such as the path basis in an interferometer). Subsequently, the properties of the density matrix are necessary to derive an upper bound for the wave-like measure in terms of the diagonal elements of the density matrix which correspond to probabilities. These properties are a consequence of the states and measurement postulates of Quantum Mechanics. In our example, the property $\rho\ge 0 \Rightarrow |\rho_{jk}|^2\le \rho_{jj}\rho_{kk}\ \forall j,k$ implies that
\begin{equation}
    C_{l_{1}}(\rho_t) \le \sum_{j\ne k}\sqrt{\rho_{jj}\rho_{kk}} \le d - 1. \label{eq:cl1}
\end{equation}
This, in turn, facilitates the identification of the corresponding predictability measure, since it has to be a function only of the diagonal elements of the density operator~\cite{Durr2001}. Hence, the inequality~\eqref{eq:cl1} suggests the following predictability measure: $P_{l_{1}}(\rho_t) = d-1-\sum_{j\ne k}\sqrt{\rho_{jj}\rho_{kk}}$. The final step involves determining the maximum value for both measures, resulting in the desired quantum complementarity inequality, which is expressed by
\begin{align}
    C_{l_{1}}(\rho_t) + P_{l_{1}}(\rho_t) \le d-1. \label{eq:cr_rho}
\end{align}

It is noteworthy that the quantum complementarity relations obtained in Ref.~\cite{Basso2020} only saturate when applied to pure quantum states, i.e., in the case of a mixed quantum state described by the quantum state preparation $\rho_t$, the quantum complementarity relations do not reach their upper bound. Notwithstanding, the way to achieve the saturation of a quantum complementarity relation is to consider that the quantum system of interest, denoted as $A$, is part of a bipartite pure quantum system $AB$ such that $\rho_t = \Tr_B (\ket{\Psi}_{AB}\bra{\Psi})$, where $\ket{\Psi}_{AB}$ serve as a purification of~$\rho_t$. Taking into account the entanglement between $A$ and $B$, inequalities in quantum complementarity relations like Eq.~\eqref{eq:cr_rho} can be transformed into equalities \cite{Basso2020_2,Basso2022}. These special quantum complementarity relations are known as quantum complete complementarity relations or quantum triality relations~\cite{Jakob2010,Jakob2012,Roy2022}.

In particular, a compelling theorem~\cite{Basso2022} was proved affirming that it is feasible to construct an entanglement monotone for a bipartite pure quantum system from any quantum complementarity relation of the form $\mf{W}(\rho_t) + \mf{P}(\rho_t) \le \alpha(d)$, which saturates only if $\rho_t$ is pure and if the wave-like measure $\mf{W}$ and the predictability measure $\mf{P}$ satisfy the required properties established in Refs.~\cite{Durr2001,Englert2008}. Then, the entanglement monotone has the form $\mf{E}(\rho_t)~=~\alpha(d)~-~\mf{W}(\rho_t) - \mf{P}(\rho_t)$. This implies that it is always possible to complete such quantum complementarity relation and, for instance, Eq.~\eqref{eq:cr_rho} can be transformed into an equality by introducing the entanglement monotone $E_{l_1}(\rho_t)$, expressed by
\begin{align}
C_{l_{1}}(\rho_t) + P_{l_{1}}(\rho_t) + E_{l_1}(\rho_t) = d - 1, \label{eq:ccrvn}    
\end{align}
where $E_{l_1}$ is equal to the robustness of entanglement for global pure states~\cite{Vidal99}. Moreover, the combination of the predictability measure with the corresponding entanglement monotone can be regarded as a path distinguishability measure \cite{Qureshi2021,Maziero2021}, effectively serving as a particle-like measure determined by the \textit{a priori} accessible information and entanglement with a path-marker~\cite{Angelo2015}. It was demonstrated that an entanglement monotone can be constructed from the largest value of distinguishability and its corresponding predictability, provided that the predictability satisfies the specified criteria \cite{Maziero2021}. This result effectively establishes a formal connection among various types of quantum complementarity relations found in the existing literature. 

In addition, it is worth mentioning that quantum complete complementarity relations of type~\eqref{eq:ccrvn} exhibit invariance under global unitary operations~\cite{Angelo2015,Basso2021,Basso2021_3}, which have important implications. Namely: (i) It allows for the consistent tracking of the wave and particle aspects of the quanton at every time instant since the quantum complete complementarity relations remain invariant under unitary evolution; (ii) It maintains its invariance under transformations of inertial reference frames in Minkowski spacetimes; (iii) It retains its invariance under transformations of local reference frames in curved spacetimes.

Grounded in these recent developments, we present an important modification of Bohr's complementarity principle, which is suggested by the connection between complementarity relations and the axioms of quantum mechanics, specifically tied to the quantum state of the system. The updated version of the quantum complementarity principle that we propose states the following: \textit{For a given quantum state preparation $\rho_t$ at a specific time instant $t$, the wave and particle behaviors of a quanton are constrained by a quantum complementarity relation $\mf{W}(\rho_t) + \mf{P}(\rho_t) \le \alpha(d)$, which is derived directly from the axioms of quantum mechanics.} 

Let us notice that our proposal introduces two main benefits. First, the quantum complementarity principle becomes an integral and inseparable component of quantum mechanics, rather than a peripheral concept existing alongside the theory. 
So, in a sense, we realize Bohr's program, once Bohr stated that the complementarity principle is a consequence of the quantum postulate but he did not actually connect the two mathematically \cite{Bohr1928,Bohr1958,Plotnitsky,khrennikov}.
Let us observe that Bohr's complementarity and Heisenberg's uncertainty principles are in the genesis of quantum mechanics. While the uncertainty principle was formulated quantitatively since the beginning of quantum mechanics, which are known as the quantum uncertainty relations and it can be derived directly from the mathematical structure of the theory \cite{Busch2007,Ozawa2014,Englert2024}, we can safely say that the quantum complementarity principle was quantitatively formulated just recently through the quantum complementarity relations. This delay in the quantitative formulation of the quantum complementarity principle produced an ongoing debate whether the uncertainty principle was the mechanism responsible for enforcing complementarity~\cite{Scully1991, Storey1994, Englert1995, Durr1998, Durr2000, Basso2021_2}. 

Second, our proposal effectively addresses potential violations of Bohr's complementarity principle and the associated complementarity relations and resolves peculiar situations reported in the literature~\cite{Wheeler1984,Ma2016, Afshar2005,Afshar2007}, thus ensuring the internal consistency of the quantum complementarity principle. Hence, in the next section, we delve into a discussion of some of these scenarios. 

%
%
\begin{figure}[t]
    \centering
    \includegraphics[scale=0.8]{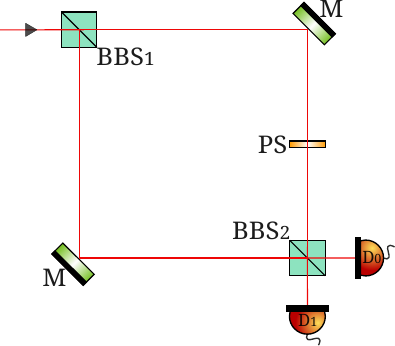}
    \caption{Variant of the Mach--Zehnder interferometer (MZI), referred to as the biased MZI (BMZI) in Ref.~\cite{Starke2023}, from which the figure was adapted. In contrast to traditional MZI, where balanced beam splitters (BS$_k$) with fixed transmission ($T_k$) and reflection ($R_k$) coefficients equal to $2^{-1/2}$, the BMZI introduces the flexibility to manipulate the amplitudes $T_k$ and $R_k$ for $k=1,2$, such that $T_k, R_k\in [0,1]$ and $T_k^2 + R_k^2 = 1$. This distinction opens avenues for enhanced control and tunability within the interferometric system. The biased beam splitter is represented by BBS$_k$, M stands for the mirrors, PS is the phase-shifter, and D$_l$ are the detectors, with $l=0,1$, for the horizontal and vertical spatial modes, respectively. 
    Defining $|0\rangle$ and $|1\rangle$ as horizontal and vertical spatial modes of the quanton, respectively, the initial state is $|\psi_0\rangle = |0\rangle$.
    The action of BBS$_k$ in $|0\rangle$ and $|1\rangle$ leads to $T_k|0\rangle + iR_k|1\rangle$ and $iR_k|0\rangle + T_k|1\rangle$, respectively. The mirrors reflect the path inserting a phase $e^{i\pi/2}=i$. The PS controls the phase difference $e^{i\phi}$ between the upper and lower paths within the BMZI. 
    }
    \label{fig:BMZI}
\end{figure}

\section{Interferometric experiments in the light of the updated quantum complementarity principle}
\label{sec:intexp}

\subsection{Biased Mach--Zenhder interferometer}
\label{sec:BMZI}

We initiate our discussion with the biased Mach--Zehnder interferometer (BMZI) introduced in Ref.~\cite{Starke2023} and illustrated in Fig.~\ref{fig:BMZI},  for which it is possible to control both transmission ($T_k$) and reflection ($R_k$) coefficients of the biased beam splitter (BBS$_k$), with $k=1,2$. The action of the BBS on the path basis is $U_{\text{BBS}}|0\rangle = T|0\rangle + iR|1\rangle$ and $U_{\text{BBS}}|1\rangle = T|1\rangle + iR|0\rangle$.

Let us then follow the evolution of the initial state $\left\vert \psi_{0}\right\rangle = \left\vert 0\right\rangle$ through the different parts of the BMZI. After passing through BBS$_{1}$, the state transforms to
\begin{equation}
\left\vert \psi_{1}\right\rangle = T_{1}\left\vert 0\right\rangle + iR_{1}\left\vert 1\right\rangle. \label{eq:psi1}
\end{equation}
The reflection due to the mirrors is described by the unitary operation $U_{\text{M}}|0\rangle=i|1\rangle$ and $U_{\text{M}}|1\rangle=i|0\rangle$, while the action of the phase shifter (PS), placed on the vertical path, is described by $U_{\text{PS}}|0\rangle=|0\rangle$ and $U_{\text{PS}}|1\rangle=e^{i\phi}|1\rangle$. Therefore, the state transforms into
\begin{equation}
\left\vert \psi_{2}\right\rangle = ie^{i\phi}T_{1}\left\vert 1\right\rangle - R_{1}\left\vert 0\right\rangle. \label{eq:psi2}
\end{equation}
Finally, after BBS$_{2}$ the state is given by
\begin{equation}
\left\vert \psi_{3}\right\rangle =  -\left( e^{i\phi}T_{1}R_{2}+R_{1}T_{2}\right) \left\vert 0\right\rangle + i\left( e^{i\phi}T_{1}T_{2}-R_{1}R_{2}\right) \left\vert 1\right\rangle. \label{eq:psi3}
\end{equation}

The interferometric visibility $\mathcal{V}$ relies on the variation of the phase introduced by the phase shifter to discern the maximum and minimum detection probabilities at D$_{l}$ (with $l=0,1$). These probabilities, derived from Born's rule for the state given by Eq.~\eqref{eq:psi3}, are expressed as
\begin{align}
\Pr\left( \text{D}_{0}\right) &= T_{1}^{2}R_{2}^{2}+R_{1}^{2}T_{2}^{2}+2T_{1}R_{1}T_{2}R_{2}\cos\phi,\\
\Pr\left( \text{D}_{1}\right) &= T_{1}^{2}T_{2}^{2}+R_{1}^{2}R_{2}^{2}-2T_{1}R_{1}T_{2}R_{2}\cos\phi,
\end{align}
which in turn leads to two distinct expressions for the interferometric visibility, i.e.,
\begin{align}
\mathcal{V}_{0} &= \frac{2T_{1}R_{1}T_{2}R_{2}}{T_{1}^{2}R_{2}^{2}+R_{1}^{2}T_{2}^{2}},\\
\mathcal{V}_{1} &= \frac{2T_{1}R_{1}T_{2}R_{2}}{T_{1}^{2}T_{2}^{2}+R_{1}^{2}R_{2}^{2}}.
\end{align}

In the particular case where BBS$_2$ is balanced, i.e., $T_2=R_2=1/\sqrt{2}$, both expressions for the visibilities coincide and are equal to the $l_1$-norm quantum coherence. However, certain scenarios emerge in which the visibility measure is inappropriate for effectively quantifying the wave-like aspect of the quanton within the interferometer. An illustrative example occurs when $R_1 = R_2 = R$ and $T_1 = T_2 = T$, resulting in $\mathcal{V}_0  = 1$ regardless of the specific values of $T$ and $R$~\cite{Starke2023}. In addition, when $T$ and $R$ take different values, the predictability measure introduced by Greenberger and Yasin \cite{Greenberger1988} yields $\mathcal{P} > 0$, thereby violating Eq.~\eqref{eq:cr} and suggesting a potential breach of the complementarity principle. However, it is crucial to note that in such cases the apparent violation arises from the inappropriate application of the wave-like measure.

Contrary to the use of visibility, where the wave behavior is inferred retroactively, the updated quantum complementarity principle relies on the use of the density matrix at a specific moment in time within the interferometer. In the case of the BMZI, any state inside the setup can be chosen for the application of the quantum complementarity relations that depend on the quantum state preparation. For instance, considering the state given by Eq.~\eqref{eq:psi2}, the corresponding density operator is given by $\rho_{2} = \left\vert \psi_{2}\right\rangle \left\langle \psi_{2}\right\vert$, which enables the evaluation of the quantum coherence and the predictability in Eq.~\eqref{eq:cr_rho}, i.e., 
\begin{align}
& C_{l_1}(\rho_{2}) = 2 T_1R_1, \label{seq:c_re} \\
& P_{l_1}(\rho_{2}) = 1 - 2T_1R_1. \label{seq:p_vn}
\end{align}
Here, it is important to note that $C_{l_1}\left( \rho_{2}\right) +P_{l_1}\left( \rho_{2}\right) =1$, since the state is pure. Hence, we are quantifying the quantum complementarity principle at the exact location where the phenomenon of interest takes place. Consequently, potential issues caused by retrodiction are circumvented, providing a more streamlined approach to assessing the complementarity behavior of a quanton within the BMZI.

Considering the insights we have just explored, it seems that the conventional method of quantifying the wave behavior based on the sensitivity of the detection probabilities to phase differences does not apply universally. Another strategy to witness the delocalized wave-like behavior of a quanton is through the creation of entanglement between interacting systems~\cite{Horodecki2009,Angelo2015}. This physically appealing method is aligned with some recent research aiming to probe the quantumness of spacetime \cite{Marletto2017,Bose2017}. In Ref.~\cite{Araujo2024}, the authors used two quantons in different arms of the interferometer, and as the degree of quantum superposition increased, the amount of entanglement between the two quantons also raised. Hence, the interferometric visibility is not the only method to measure the quantum superposition. 
Moreover, given that it is possible to represent any density operator using a basis of Hermitian operators, by determining the expectation values of these operators, the system's state can, in principle, be reconstructed through a technique called quantum state tomography \cite{James2001,Lundeen2011,Tekkadath2016}. This approach allows for the direct witness or quantification of both wave and particle behaviors from the system's density matrix. In essence, quantum state tomography offers a more versatile way of characterizing the system, bypassing the constraints associated with sensitivity to phase differences.

%
%
\subsection{Wheeler's delayed choice experiment}
\label{sec:WDCE}

We can extend our previous analysis to Wheeler's delayed choice experiment (WDCE) \cite{Wheeler1984,Ma2016} by using the same apparatus as depicted in Fig.~\ref{fig:BMZI}. 
Let us notice that, when $T_2 = 1$, the effect is the same as the absence of BBS$_2$, while, for $R_2 = T_2 = 2^{-1/2}$, the effect is the same as the presence of a balanced beam splitter, BS$_2$. Hence, when BS$_{2}$ is absent ($T_2 = 1$), the  phase sensitivity leads to a null visibility and the detectors reveal a particle-like behavior, according to Ref.~\cite{Ionicioiu2011}. However, it should be observed that we can have all sorts of quantum superposition inside the interferometer, once $T_1$ and $R_1$ are free to vary, and we still have $\mathcal{V}=0$. In particular, when $R_1 = T_1 = 2^{-1/2}$, we have the maximum superposition inside the interferometer and the visibility is null, which implies that the visibility does not reflect the quantum superposition inside the interferometer.

On the other hand, when the BS$_2$ is present, the visibility correctly quantifies the quantum superposition inside the interferometer, being equal to the $l_1$-norm quantum coherence. Moreover, the delayed aspect of this experiment implies that the observer retains the flexibility to decide until the last moment which behavior will be recorded in the detectors. In addition, Ref.~\cite{Ionicioiu2011} enhanced Wheeler's experiment by considering the second beam splitter in a superposition of being inside and outside the interferometer, the so called quantum delayed choice experiment (QDCE).

This apparent back-in-time causation is effectively avoided through our updated version of the quantum complementarity principle. Inclusion or exclusion of BS$_2$ has no impact on the wave or particle-like behavior at earlier times.  In particular, if BBS$_1$ is balanced, we have $C_{l_1}(\rho_t) = 1$ and $P_{l_1}(\rho_t) = 0$ inside the interferometer, reflecting the maximum path quantum superposition, regardless of the choice to place or not the BBS$_2$. We observe that Ref.~\cite{Dieguez2022} further explores this issue by demonstrating that the wave element of reality, as defined by the authors, is consistently maximum in the QDCE, aligning with our updated version of quantum complementarity principle. Those authors also established complementarity relations in terms of wave and particle elements of reality, which depend on the system quantum state preparation, fitting into our proposal.

%
%
\begin{figure}[t]
    \centering
    \includegraphics[scale=0.84]{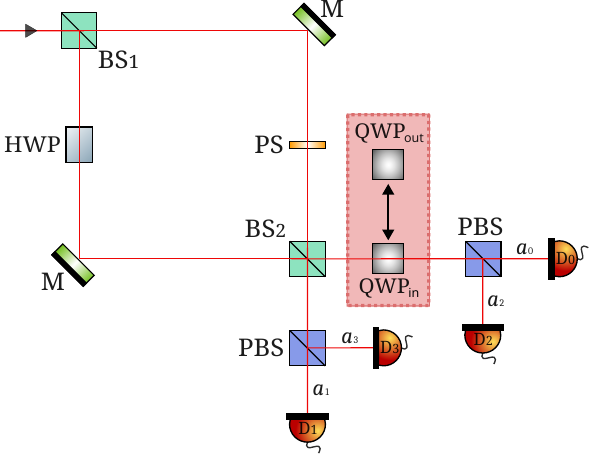}
    \caption{
    Modified representation of the partial quantum eraser (PQE) discussed on Refs.~\cite{Pessoajr2013,Pessoajr2000,Herzog1995}.
    The beam splitter is represented by BS$_m$ with $m=1,2$, M stands for the mirrors, HWP is a half-wave plate, PS is the phase-shifter, QWP is a quarter-wave plate that can be inserted into the apparatus (QWP$_{\text{in}}$) or removed (QWP$_{\text{out}}$), PBS is a polarizing beam splitter and D$_n$ are the detectors, with $n=0,1,2,3$.
    Horizontal and vertical polarization modes (a subscript represented by a capital letter $A$) are defined by $|0\rangle$ and $|1\rangle$, respectively. Similarly, the horizontal and vertical spatial modes (a subscript represented by a lowercase letter $a$) of the quanton are defined as $|0\rangle$ and $|1\rangle$, respectively.
    HWP inverts the input polarization, i.e., $U_{\text{HWP}}|0\rangle_A = |1\rangle_A$ and $U_{\text{HWP}}|1\rangle_A = |0\rangle_A$. QWP converts a linear polarization into a circular polarization, i.e,, $U_{\text{QWP}}|0\rangle_A = |\oplus\rangle_A$ and $U_{\text{QWP}}|1\rangle_A = |\ominus\rangle_A$, where $|\oplus\rangle = 2^{-1/2}(|0\rangle + i|1\rangle)$ and $|\ominus\rangle = 2^{-1/2}(|0\rangle - i|1\rangle)$.
    The PBS only transmits horizontal polarization and reflects vertical polarization, so that $U_{\text{PBS}}|00\rangle_{aA} = |00\rangle_{a_0A}$, $U_{\text{PBS}}|01\rangle_{aA} = i|11\rangle_{a_2A}$, $U_{\text{PBS}}|10\rangle_{aA} = |10\rangle_{a_1A}$ and $U_{\text{PBS}}|11\rangle_{aA} = i|01\rangle_{a_3A}$, where the subscript $a_n$ at the output of the PBS represents the path leading to the detector D$_n$.
    }
    \label{fig:local_QE}
\end{figure}

%
%
\subsection{Partial quantum eraser}
\label{sec:PQE}

The partial quantum eraser (PQE), as discussed in Refs.~\cite{Pessoajr2013,Pessoajr2000,Herzog1995} and depicted in Fig.~\ref{fig:local_QE}, presents another compelling case that highlights the necessity of our proposal. 

Let us consider that, in addition to the spatial mode, the quanton enters the MZI with horizontal polarization. The lower path is equipped with a half-wave plate (HWP), which changes the horizontal to vertical polarization. Consequently, the paths can be distinguished through the photon's polarization, and thus the paths become distinguishable. Outside the MZI, one can choose to include (QWP$_{\text{in}}$) or not (QWP$_{\text{out}}$) a quarter-wave plate (QWP), which is responsible for converting linear polarization into circular polarization. The polarized beam splitter (PBS) is designed to transmit horizontal polarization and reflect vertical polarization. Consequently, each detector will only click to a specific polarization. For QWP$_{\text{out}}$, the click on detectors D$_0$ or D$_1$ is linked to the upper path, while detection in D$_2$ or D$_3$ is linked to the lower path. On the other hand, for QWP$_{\text{in}}$, the path information becomes effectively erased in the detectors D$_0$ and D$_2$. If a detection occurs at D$_1$ or D$_3$, it is still possible to infer the path of a quanton.

Quantitatively, let us consider that the initial state is \(|\psi_0\rangle = |00\rangle_{aA}\), where the spatial mode is represented by the subscript $a$ while the polarization is represented by the subscript $A$. Horizontal and vertical polarization are represented by $|0\rangle$ and $|1\rangle$, respectively. After passing through the BS$_1$, the state becomes
\begin{equation}
|\psi_1\rangle = \frac{1}{\sqrt{2}}(|0\rangle_{a} + i|1\rangle_a)|0\rangle_A.\label{seq:lqepsi1}
\end{equation}
The action of the HWP inverts the polarization yielding
\begin{equation}
|\psi_2\rangle = \frac{1}{\sqrt{2}}(|00\rangle_{aA} + i|11\rangle_{aA}),\label{seq:lqepsi2}
\end{equation}
which implies that there is entanglement between the path and the polarization of the system.
As the system passes through the mirrors, phase-shifter, and BS$_2$, the quantum state evolves to
\begin{equation}
|\psi_3\rangle = -\frac{1}{2}[|0\rangle_{a}(e^{i\phi}|0\rangle_A + |1\rangle_A) - i|1\rangle_{a}(e^{i\phi}|0\rangle_A - |1\rangle_A)].
\end{equation}

If the quarter-wave plate is absent, i.e., for QWP$_{\text{out}}$, the state is then given by
\begin{equation}
|\psi_4^{\text{out}}\rangle = -\frac{1}{2}(e^{i\phi}|00\rangle_{a_0A} + i|11\rangle_{a_1A} - ie^{i\phi}|10\rangle_{a_2A} - |01\rangle_{a_3A}),
\end{equation}
with the transmission of the horizontal polarization and the reflection of the vertical polarization by the PBS. A direct calculation shows that this setup yields the probability of $1/4$ for all detectors, allowing for the distinction between the upper and lower paths.

On the other hand, if the quarter-wave plate is present, i.e., for QWP$_{\text{in}}$, the state of system after the PBS becomes 
\begin{align}
|\psi_4^{\text{in}}\rangle =  -\frac{1}{2\sqrt{2}}[(e^{i\phi}+1)|00\rangle_{a_0A} - (e^{i\phi}-1)|11\rangle_{a_2A}   - i\sqrt{2}e^{i\phi}|10\rangle_{a_1A} - \sqrt{2}|01\rangle_{a_3A})].
\end{align}
The probabilities for the detectors D$_0$ and D$_2$ are respectively given by
\begin{align}
\Pr\left(  D_{0}\right)   &  = \frac{1}{4}\left(  1+\cos\phi\right)  ,\\
\Pr\left(  D_{2}\right)   &  = \frac{1}{4}\left(  1-\cos\phi\right),
\end{align}
while the probabilities at D$_1$ and D$_3$ remain untouched, giving the probability of $1/4$ for each detector. Here, the probabilities at D$_0$ and D$_2$ are indeed sensitive to phase variation. Therefore, based on the definition of wave behavior through phase variation, we may conclude that the wave pattern is restored within the interferometer when detectors D$_0$ or D$_2$ clicks. Conversely, path information is obtained when detectors D$_1$ or D$_3$ clicks.

Similarly to the analysis for Wheeler's delayed choice experiment, we can use the quantum complementarity relation that depends on the quantum state in each step of the interferometer to better understand what happens in the partial quantum eraser. The state given by Eq.~\eqref{seq:lqepsi1} results in a separable state between the polarization and the path degrees of freedom with maximum path quantum superposition. Consequently, the quantum complete complementarity relations given by Eq.~\eqref{eq:ccrvn} yields $C_{l_1}(\rho_{1,a}) = 1$ and $P_{l_1}(\rho_{1,a}) = E_{l_1}(\rho_{1,a}) = 0$, where $\rho_{1,a} = \Tr_A\left(|\psi_1\rangle \langle \psi_1|\right)$. That is to say, there is no path information yet, only the wave behavior.

After the action of the HWP, the state is given by Eq.~\eqref{seq:lqepsi2}, we can see that there is entanglement between the polarization and the path degrees of freedom, which provides a path marker. The quantum complete complementarity relation given by Eq.~\eqref{eq:ccrvn} leads to $C_{l_1}(\rho_{2,a}) = P_{l_1}(\rho_{2,a}) = 0$ and $E_{l_1}(\rho_{2,a}) = 1$, where $\rho_{2,a} = \Tr_A\left(|\psi_2\rangle \langle \psi_2|\right)$.

In contrast with the biased Mach-Zehnder interferometer (see Sec.~\ref{sec:BMZI}), where the choice of state inside the interferometer does not matter for quantifying the wave and particle aspects of the system, here the choice is important once the HWP alters drastically the quantum state. In the partial quantum eraser, the quantum coherence is converted into entanglement when there is a change in the polarization by the HWP. Therefore, this setup highlights the subtlety of the updated version of the quantum complementarity principle and how the wave and particle behavior changes through the time evolution of the quantum system.

%
%
\subsection{Unruh's experiment: an analogue to the Afshar's experiment}
\label{sec:AE}

Let us now consider Afshar's experiment (AE)~\cite{Afshar2005,Afshar2007}, which is an interesting variation of Young's double-slit experiment \cite{Maleki2023}. Let us recall that Afshar claimed that, in his experimental setup, both wave and particle measures can attain their maximum values, thus challenging Bohr's complementarity principle. 

\begin{figure}[t]
\centering
\includegraphics[scale=0.94]{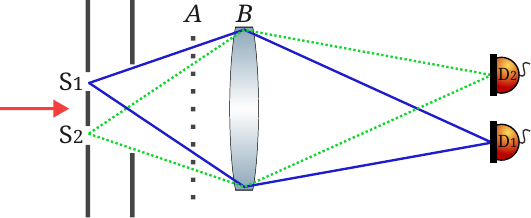}
\caption{
Schematic depiction of Afshar's experiment (AE). The incident quanton travels from left to right, passing through slits S$_1$ and S$_2$. The solid and dashed lines represent the quanton paths through slits S$_1$ and S$_2$, respectively. At location $A$, vertical dotted lines denote wires oriented perpendicularly to the page, meticulously positioned at the interference pattern minima, enabling the capture of wave-like information. At region $B$, a lens supposedly redirects the interfered quanton to detector D$_q$, where $q=1,2$.
Afshar claimed that this configuration allows for the determination of the slit from which the quanton originated. This figure was adapted from Ref. \cite{Afshar2005}.
}
\label{fig:afshar}
\end{figure}

In the experimental setup depicted in Fig.~\ref{fig:afshar}, an incident quanton travels from the left to the right and passes through the slits S$_1$ and S$_2$ and interferes maximally within region $A$. In region $A$, the wire grids, perpendicular to the plane of the page, are meticulously positioned at the interference pattern minima. These grids are designed to capture wave-like information through a non-perturbation criterion. At position $B$, a lens is employed to direct the quanton originating from the slit S$_q$ to detector D$_q$, with $q=1,2$.

The author used a different experimental configuration, in which one of the slits was closed, resulting in detection occurring exclusively in the detector corresponding to the open slit. Thus, when both slits are open, interference is observed at position $A$, and the quanton continues through the lens. Upon arrival at one of the detectors, the author states that there is a violation of Bohr's principle of complementarity, because at position $A$ we have maximum wave behavior (maximum $W$) and the click of one of the detectors will inform which slit the photon passed (maximum $P$).
Since its conception, AE has been the subject of extensive discussions among researchers and has prompted ongoing investigations aimed at resolving the apparent paradox~\cite{Qureshi2012,Steuernagel2007,Georgiev2012,Jacques2008,Knight2020,Gergely2022,Qureshi2023,Unruh2004}.

%
%
\begin{figure}[t]
    \centering
    \includegraphics[scale=0.8]{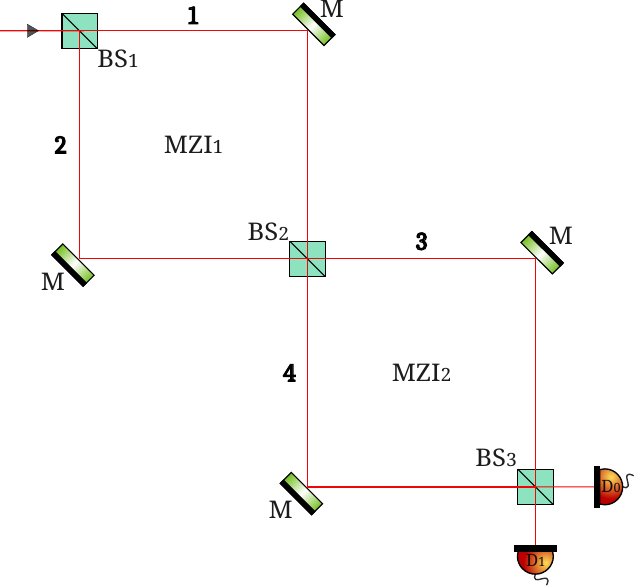}
    \caption{
    A schematic depiction of Unruh's experiment, adapted from Ref.~\cite{Unruh2004}, which involves the sequential utilization of two Mach--Zehnder interferometers (MZIs) used as an analogous case to Afshar's experiment employing two-level systems. After traversing MZI$_1$, the quantum state achieves maximum path coherence. Blocking either path 1 or path 2 results in detections exclusively at detector D$_0$ or D$_1$, respectively. By analogy with Afshar's reasoning, when both paths 1 and 2 are unobstructed, it becomes feasible to deduce the path taken by the quanton within MZI$_1$ through measurements at the detectors. Importantly, this deduction occurs while maintaining maximum wave-like behavior, as per the reasoning outlined by Afshar.
    }
    \label{fig:unruh}
\end{figure}

To elucidate the concepts discussed in the AE, we can examine a simpler but analogous experiment proposed by Unruh that involves a two-level system \cite{Unruh2004}. The experiment features two MZIs arranged in sequence, as illustrated in Fig.~\ref{fig:unruh}. The paths of the MZI$_1$ play the role of the two slits in the AE, while MZI$_2$ mimics the region of the wire grid in the AE.
By obstructing the path {\bf 2}, all quantons passing through the path {\bf 1} are directed to detector D$_1$. Conversely, blocking path {\bf 1} results in all quantons from path {\bf 2} being measured at D$_0$. When both paths in MZI$_1$ are unblocked, we will have constructive (destructive) interference in path {\bf 3} ({\bf 4}) of the MZI$_2$. Therefore, the conclusions drawn from the analysis of the Unruh's experiment can also be used in the Afshar experiment and vice versa.

Let us then follow the evolution of the quantum state in Unruh's experiment. As depicted in Fig.~\ref{fig:unruh}, the initial state is given by
\begin{equation}
    \left\vert \psi_{0}\right\rangle =\left\vert
0\right\rangle. \label{seq:ui}
\end{equation}
The first beam-splitter creates a maximum quantum superposition such that the state right after BS$_1$ is given by
\begin{align}
    \left\vert \psi_{1}\right\rangle =\frac{1}{\sqrt{2}%
}\left(  \left\vert 0\right\rangle+i\left\vert 1\right\rangle
\right), \label{eq:u1}
\end{align}
while inside the second MZI the state is given by
\begin{equation}
    \left\vert \psi_{2}\right\rangle 
=-\left\vert 0\right\rangle.\label{eq:u3}
\end{equation}
Finally, after the third beam-splitter, the state of the system becomes
\begin{align}
    \left\vert \psi_{3}\right\rangle =\frac{1}{\sqrt{2}%
}\left(  \left\vert 0\right\rangle-i\left\vert 1\right\rangle
\right). 
\end{align}
In the AE, this situation corresponds to the ordinary situation of two unblocked slits. Now, if we block the path {\bf 2} and consider the same initial state given by Eq.~\eqref{seq:ui}, all quantons that pass through the path {\bf 1} will be described by the normalized state
\begin{align}
    \left\vert \psi_{1}^{\prime
}\right\rangle =\left\vert 0\right\rangle, \label{eq:u1'}
\end{align}
which implies that the state right after the second beam splitter will present maximum quantum superposition, i.e., 
\begin{align}
    \left\vert \psi_{2}^{\prime}\right\rangle =-\frac
{1}{\sqrt{2}}\left(  \left\vert 0\right\rangle-i\left\vert 1\right\rangle
\right), \label{eq:u3'}
\end{align}
which implies that, according to Afshar, the upper path in MZI$_1$ will always be identified with the detections in D$_1$, once the state after the third beam splitter is given by
\begin{align}
    \left\vert \psi_{3}^{\prime}\right\rangle
=-i\left\vert 1\right\rangle.
\end{align}

Similarly, if we block the path {\bf 1}, all quantons that pass through the path {\bf 2} will be described by the normalized state $\left\vert \psi_{1}^{\prime
\prime}\right\rangle =i\left\vert 1\right\rangle $. Once again, the state right after the second beam splitter will present maximum quantum superposition and, therefore, the lower path in MZI$_1$ will always be identified by the detections in D$_0$, given that the state after the third beam splitter becomes $\left\vert \psi_{3}^{\prime\prime}\right\rangle
=\left\vert 0\right\rangle.$ Hence, each unblocked path is associated with one detector.

Now, let us analyze Unruh's experiment in light of Afshar's argumentation. When both paths in MZI$_1$ are unblocked, which corresponds to the quanton coming from both slits in the AE, we will have constructive interference in the path {\bf 3} of MZI$_2$ and destructive interference in the path {\bf 4} of MZI$_2$, which corresponds to the region of the wires grid in the AE. After this, Afshar argues that a detection at D$_0$ or D$_1$ will reveal which path the quanton took before being detected. Therefore, Afshar concludes that we have a violation of the Bohr's complementarity principle, since we have maximum wave-like behavior and maximum particle-like behavior in the same experimental setup.

Here, one can see that two different assumptions are made regarding these experiments. The first assumption is the acceptance of quantifying Bohr's complementarity principle in the whole experimental setup with the wave and particle behavior being quantified in different parts of the experiment. In the case of the Afshar experiment, the retrodiction occurs in the quantification of the particle behavior, which associates the detection in one of the detectors with one of the slits, while the wave-like behavior is quantified in the region of the wires grid by the interferometric visibility. Beyond that, in the experiment reported in Ref.~\cite{Afshar2007}, in order to calculate the wave and particle measures, it was necessary to have three different configurations: no wire grid, wire grid in central minima, and wire grid in central minima and one pinhole blocked. 

The second assumption is that the conclusions drawn about one experimental setup remain valid in the other experimental setup. More specifically, the conclusion that the click of each detector corresponds to a particular path which the quanton went through, when one of the path is blocked, remains valid when none of the paths are blocked. These situations (one path blocked and both paths open) correspond to different experimental setups (or different configurations of the same experimental setup), which was already criticized in Ref.~\cite{Unruh2004}.

From the perspective of the updated quantum complementarity principle, the question becomes clear.  When both paths are open, the state presents maximum path quantum superposition in the MZI$_1$ region, while it presents maximum path predictability in MZI$_2$. In contrast, when one of the paths is blocked, the state presents maximum path predictability in the MZI$_1$ region, while it presents the maximum quantum superposition of the path in MZI$_2$. Therefore, it is easy to see that the quantum complementarity relation~\eqref{eq:cr_rho} is always satisfied in each step of the different experimental setups.

%
%
\section{Final remarks}
\label{sec:conc}

Although Heisenberg's uncertainty principle was deeply rooted in the foundations of quantum mechanics from its early days, Bohr's complementarity principle encountered a more enigmatic genesis. By focusing on the whole experimental setup, conceptual challenges were raised, historically exemplified by the Wheeler's delayed choice experiment and the Afshar's experiment. Our proposal sought to address these challenges by advocating an updated quantum complementarity principle, grounded in the complementarity relations derived directly from the postulates of quantum mechanics. These quantum complementarity relations, emerging from the quantum postulates, highlights the role of the density matrix at each moment in time. 

In the Wheeler's delayed choice experiment, we showed that retroactive causation is effectively circumvented given that the inclusion or not of the second beam splitter does not influence the wave or particle-like behavior observed at earlier moments. Regarding the Unruh's version of the Afshar experiment, when both paths are available, the state exhibits maximum path quantum superposition in MZI$_1$ and maximum path predictability in MZI$_2$. Conversely, if one of the paths is obstructed, the state shows maximum path predictability in the MZI$_1$ region and maximum quantum superposition in MZI$_2$, without the need of retroactive inference about the particle-like behavior. 

Moreover, we also discussed two other experimental setups that highlights the subtlety of our proposal. The biased Mach-Zehnder interferometer display situations in which the interferometric visibility does not quantifies the wave-like behavior correctly and thus the quantification of the wave and particle aspect through the quantum complementarity relations, that depends on the system's density matrix, is more appropriate. The partial quantum eraser demonstrates that the quantum coherence turns into entanglement, which implies that the wave and particle-like behaviors changes within the interferometer.

So, in this article, we presented a novel approach to the quantum complementarity principle. This principle holds as long as quantum mechanics applies. Additionally, our new formulation sheds light in experiments that remained veiled in conceptual ambiguity. We believe that this refinement of the quantum complementarity principle will be of fundamental importance for the foundational basis of quantum theory.

\vspace*{0.5cm}
\begin{acknowledgments}
\vspace*{-0.4cm}
This work was supported by the 
Coordination for the Improvement of Higher Education Personnel (CAPES),
Grant No.~88887.827989/2023-00,
the S\~ao Paulo Research Foundation (FAPESP),
Grant No.~2022/09496-8.3,
the National Council for Scientific and Technological Development (CNPq),
Grants No.~309862/2021-3,
No.~409673/2022-6, and
No.~421792/2022-1,
and the National Institute for the Science and Technology of Quantum Information (INCT-IQ), 
Grant No.~465469/2014-0.
\end{acknowledgments}


\end{document}